\documentclass[runningheads]{llncs}
\usepackage[T1]{fontenc}
\usepackage{graphicx}
\usepackage{subcaption}
\usepackage{booktabs} 
\usepackage{hyperref}
\begin{document}
\title{Enhancing the automatic segmentation and analysis of 3D liver vasculature models}
\titlerunning{Paper accepted at MICCAI 2024 Workshop: ADSMI}
\author{Yassine Machta\inst{1} \and
Omar Ali\inst{1,2}\and
Kevin Hakkakian\inst{1,3}\and
Ana Vlasceanu\inst{1,4}\and
Amaury Facque\inst{1,3}\and
Nicolas Golse\inst{3} \and
Irene Vignon-Clementel\inst{1}}
%
\authorrunning{Machta et al.}
%
\institute{INRIA, Palaiseau \and
Guerbet, Paris \and
AP-HP, Université Paris-Saclay and Inserm U1993 \and CHU Tours}

%
\maketitle     

\begin{abstract}
Surgical assessment of liver cancer patients requires identification of the vessel trees from medical images. Specifically, the venous trees - the portal (perfusing) and the hepatic (draining) trees are important for understanding the liver anatomy and disease state, and perform surgery planning.
This research aims to improve the 3D segmentation, skeletonization, and subsequent analysis of vessel trees, by creating an  automatic pipeline based on deep learning and image processing techniques.

The first part of this work explores the impact of differentiable skeletonization methods such as ClDice and morphological skeletonization loss, on the overall liver vessel segmentation performance. To this aim, it studies how to improve vessel tree connectivity.

The second part of this study converts a single class vessel segmentation into multi-class ones, separating the two venous trees. It builds on the previous two-class vessel segmentation model, which vessel tree outputs might be entangled, and on connected components and skeleton analyses of the trees.

After providing sub-labeling of the specific anatomical branches of each venous tree, these algorithms also enable a morphometric analysis of the vessel trees by extracting various geometrical markers.

In conclusion, we propose a method that successfully improves current skeletonization methods, for extensive vascular trees that contain vessels of different calibers. The separation algorithm creates a clean multi-class segmentation of the vessels, validated by surgeons to provide low error. A new, publicly shared high-quality liver vessel dataset of 77 cases is thus created. Finally a method to annotate vessel trees according to anatomy is provided, enabling a unique liver vessel morphometry analysis.

\keywords{3D liver vessel segmentation  \and Morphometry Analysis \and Automatic labeling \and Liver Dataset \and Liver vessel label propagation}
\end{abstract}
%
%
\section{Introduction}
Having accurate 3D geometrical models of liver vessels is an important step to plan surgical liver resection \cite{core}, the only curative treatment of liver cancer, which continues to be a growing healthcare problem worldwide \cite{rumgay2022global}. Such models also allow to study the morphometry of different vascular trees in healthy or diseased livers. To obtain these models, achieving reliable 3D segmentation is a crucial step. However, deep-learning (DL) based vessel segmentation suffers from drawbacks such as disconnections or label mix-ups that have been studied with graph-based approaches \cite{scope}, persistent-homology approaches \cite{ph2}, and vessel contrast filters \cite{Jerman}  approaches featuring multiple decoders \cite{keshwani}.
%

The first part of this research builds on the study of skeleton-based topological loss functions. These functions introduced as ClDice by \cite{cldice}, offer a way to improve vessel continuity in the segmented volume. The Cl-Dice method leverages a differentiable algorithm to soft-skeletonize a tubular structure in order to supervise the training with this skeleton's topological information. MS Loss \cite{msloss} presents the  Soft-Persistent-Skeletonization. It is also a differentiable method of skeletonization that preserves connectivity. However these two methods may yield to sub-optimal results, featuring many disconnections and uneven thickness of the centerline. The reasons are the segmentation's low voxel resolution (512x512x~512 voxels compared  to the kernel of the morphological operation) and the irregular tree diameters.
Alternative methods utilize gradient based optimisation \cite{menten} or the Unet to obtain the skeletonization in 2D \cite{nnskel}. In this paper, we investigate which skeletonization method is best for liver vessel trees, and whether an improved vessel skeleton can improve the results of 3D vessel segmentation.

A major limitation in DL liver vessel segmentation is the lack of public, annotated with high-quality, and large datasets, which on top distinguish between the portal and hepatic trees. Hence, we also evaluate the impact of increasing the size of the dataset on the overall liver vessel segmentation performance. To that end, we create a tool that can alleviate the annotation burden of vessel volumes. It automatically converts a single-class annotation of the vessels into a multi-class version separating the portal and hepatic trees. It uses heuristics to assign conflicted parts of the voxel skeleton as opposed to the probabilistic model by \cite{kang2014automatic} used on meshes. This algorithm is run on an in-house-annotated subset of the public LiTS \cite{lits} dataset, to obtain new pairs of CT scan and vessel annotations featuring the portal and hepatic trees alongside the already available liver and tumor segmentations. This new dataset is published with this paper.

The final part of the pipeline leverages the segmentation obtained above to provide automatic analysis, relevant for pathophysiological and surgical applications \cite{portal}. We propose a novel method to further label individual branches, building on an improved automatic segmentation of the different clinical segments of the liver volume as defined by the surgical Couinaud system \cite{couinaud}. This enables a statistical analysis of the branches and patterns of both venous trees, that provides morphometric information valuable not only for a general understanding of vascularized organs but also for disease staging and liver regeneration understanding.

\subsubsection{Contributions.}
In this paper we evaluate  the impact of an alternative differentiable skeletonization method on the performance of the ClDice loss in a 3D setting for liver vessel segmentation, and provide 
a tool to automatically separate the different vessel trees  of vascularized organs, applied and tested on the liver vessel trees in this research. 
We publicly release the dataset obtained from these experiments after validation from surgeons.
Finally, to the best of our knowledge, we also provide a method to automatically and anatomically further label the different tree branches down to the bifurcation level, along with a morphometric analysis of these labeled liver portal and hepatic trees, performed on the largest public liver vessel dataset to date.


\section{Methods}
\subsection{Deep learning techniques to improve vessel tree skeletonization and segmentation}

\subsubsection{Skeleton based losses and differentiable skeletonization methods.}
\label{subsec:model}
\begin{figure}
    \centering
    \caption{Model description for multi-class vessel segmentation (right box) which uses the pretrained skeletonization network: NeuralSkel in its loss (left box). GT= ground truth.}
    \label{fig:model}
    \includegraphics[width=1.0\linewidth]{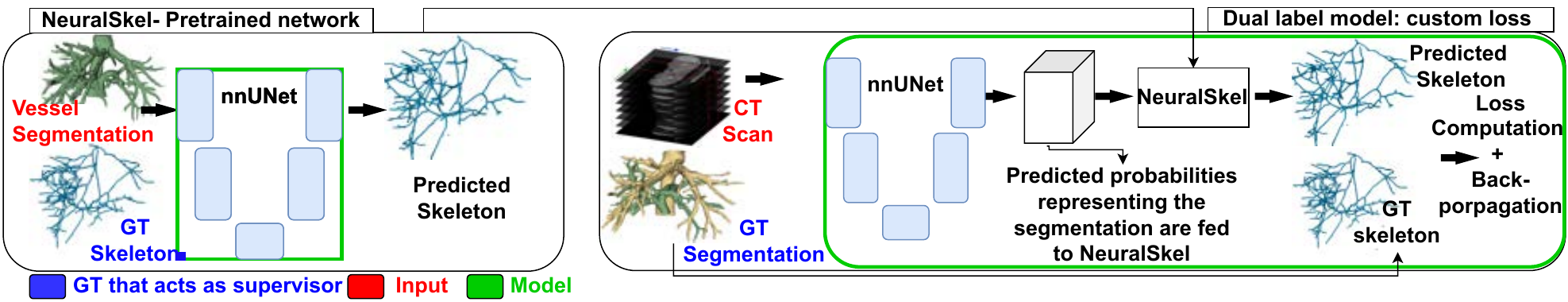}
    
\end{figure}

Building on the works of Nguyen et al. \cite{nnskel}, an nnunet \cite{nnunet} was pretrained to predict the liver vessel centerlines  as shown on the right in Fig.\ref{fig:model}. The ground truth (GT) leveraged for the  training of the network is the centerline of the different liver vessels obtained with Lee's \cite{lee} skeletonization method implemented in the Scikit-image library. This skeletonization neural network is then deployed on the softmax outputs of the vessel segmentation model before being leveraged for the computation of the ClDice loss. (Fig.\ref{fig:model})



\subsubsection{Model properties, hardware and datasets.}
The different experiments (\ref{fig:model}) are performed using the nnUNet framework \cite{nnunet}, with a modified loss incorporating ClDice in addition to the  default dice and cross entropy losses. Training processes take 14 to 20 hours for 500 epochs per fold for 5 folds. The multi-class liver vessel dataset, encompassing both portal and hepatic trees, originates from the public IRCAD dataset, which consists of 20 CT volumes  \cite{ircad}, split into 16/4 training/validation volumes respectively. To palliate the small size dataset, we were able to add 53 cases segmented from the public LiTS \cite{lits} dataset (see the next section \ref{sec:separation}). Overall, the model was trained on 69 cases and validated on the same set of 4 volumes from IRCAD for all the folds .

\subsection{Algorithm to separate portal and hepatic venous trees}
\label{sec:separation}

\begin{figure}
    \centering
    \caption{Vascular tree separation algorithm; CC: Connected component.}
    \label{fig:separation}
    \includegraphics[width=1.0\linewidth]{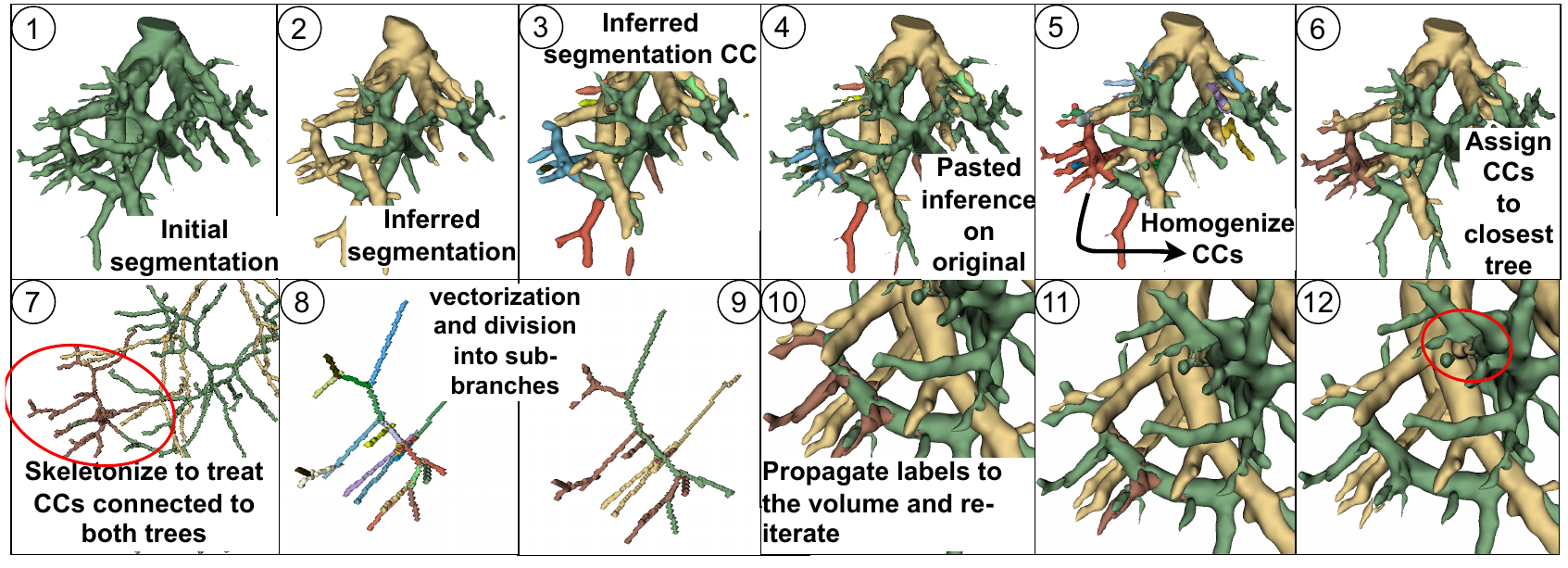}

\end{figure}

We propose an algorithm that, from a single class venous vessel segmentation, automatically distinguishes between the portal and hepatic vessel trees. During the whole process described in Fig.\ref{fig:separation}, the original segmentation remains intact, where only the branches belonging to their respective trees are labeled accordingly. Numbers in \textit{italic} refer to steps in Fig.\ref{fig:separation}, that are described below. 

We begin by "pasting" an inferred dual labeled segmentation, thanks to the model trained on IRCAD in section \ref{subsec:model}, on the original single-label one from our inhouse annotation of LiTS vessels (\textit{2}). We do this without compromising the original segmentation. Due to a general phenomenon of under segmentation by the model, we dilate the inferred segmentation.

The uncertain parts that were not covered by the inference are treated next (\textit{3-5}). Our course of action is to isolate these regions by finding the different connected components (CC). They are then assigned to the closest main tree (\textit{6}). We also filter out certain classes that are farther than both trees by a fixed distance or too small (<50 voxels), to avoid floating patches in the segmentation.

If a conflict class is in contact with both portal and hepatic trees, we proceed to skeletonize the structure. The skeleton is turned into branches according to bifurcations, by the skan package \cite{skan}. Once we have separated and labeled the branches, we transform them into vectors by creating a Bersenham line between the first and last point of the branch (\textit{7-8}).


We iterate through all the branches starting by the closest ones to the portal tree input(s) and the hepatic tree input(s). We determine whether it is connected if the angle between the branches is inferior to 60° \cite{murray} (\textit{9}).

This process of CCs and skeleton analysis is re-iterated until the disappearance of conflicted classes(\textit{10-12}). If there is no progress, the algorithm stops.


\subsection{Tree automatic anatomical labeling and morphological analysis}

We first extract the 1-pixel wide centerline of the tree with Lee's \cite{lee} skeletonization method. The rest of the analysis is completed with the libraries Sknw (https://github.com/Image-Py/sknw) and NetworkX, following a method by Decrooq et al. \cite{decroocq2022software}. 

\textbf{Radius and Length} are obtained by further exploring the networkX graph obtained from the skeleton. Since the original volumes are voxel based, graphs also have integer coordinates. But taking into account the voxel dimensions (or the sampling of the CT Scans) leads to an accurate metric representation of the tree: the length and radii of each branch are computed after a simple graph search. Here, we employ a Breadth first search after automatically finding the root of the two trees.

\textbf{Vascular tree roots} 
The liver segmentation enables the detection of the roots as they are the farthest tree endpoints from the liver volume: the roots are outside the liver.

\textbf{Branching angles properties}
 Due to noise created by the skeletonization and sknw library, this process is not trivial as it can create extremely short paths which are not relevant, or loops which are unrealistic.
But they can be fused to obtain an oriented graph, starting from the real anatomical root of the tree. We can now analyze the morphometry of the trees. 

\textbf{Automatic labeling according to anatomical terminology}
In order to visualize and analyze the morphometric data, we need a method to label and locate each branch. There exist many indexing methods such as Strahler's order \cite{Strahler}. Here, we settled on the anatomical nomenclature commonly referred to by liver surgeons and clinicians. This requires the ability to pinpoint, for example for the portal vein, the main portal vein, right branch, left branch, the different anterior or posterior segments. This is performed by heuristics relying on branch relations and spatial position. But we also leverage the fact that since liver vessels enable the classification of the liver volume into anatomical segments called the Couinaud segments. The prediction of these segments by a plain nnUNet model trained on the dataset provided by Tian et al. \cite{couinaud}, is added to help in anatomically distinguishing between branches. To pinpoint the branch root of the portal tree for example, we keep fusing paths until the descendants of the main tree are in separate parts of the liver, right and left (Fig. \ref{fig:annotation}c).

\begin{figure}
    \centering
    \caption{Vessel tree skeletonization methods}
    \label{fig:skel}
    \begin{tabular}{cccc}
        \includegraphics[width=0.23\linewidth]{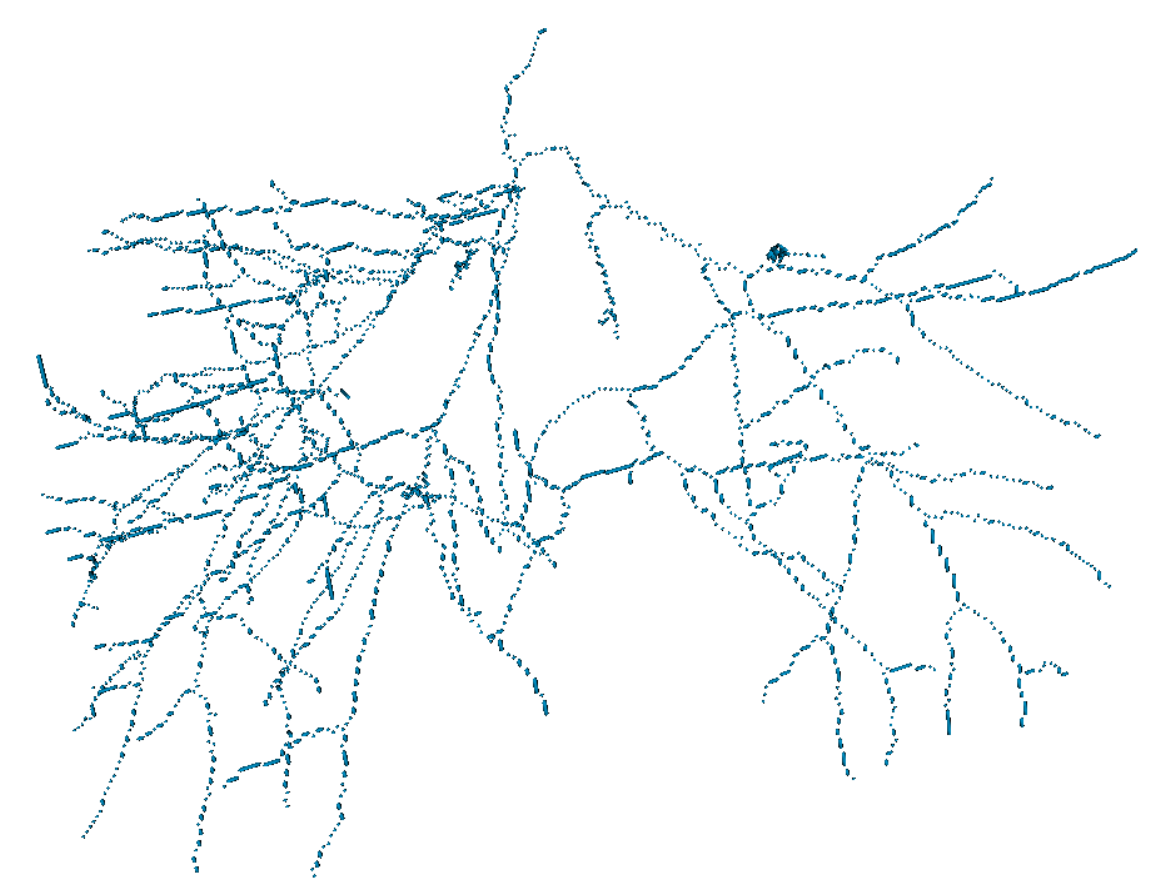} &
        \includegraphics[width=0.23\linewidth]{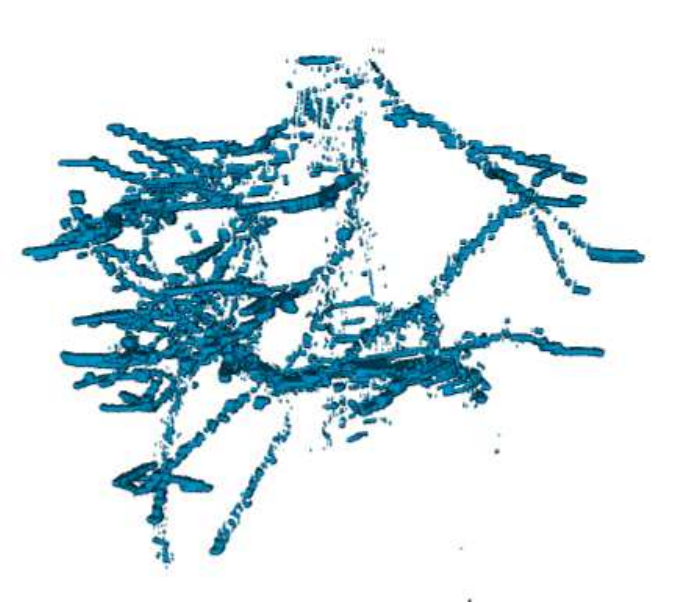} &
        \includegraphics[width=0.23\linewidth]{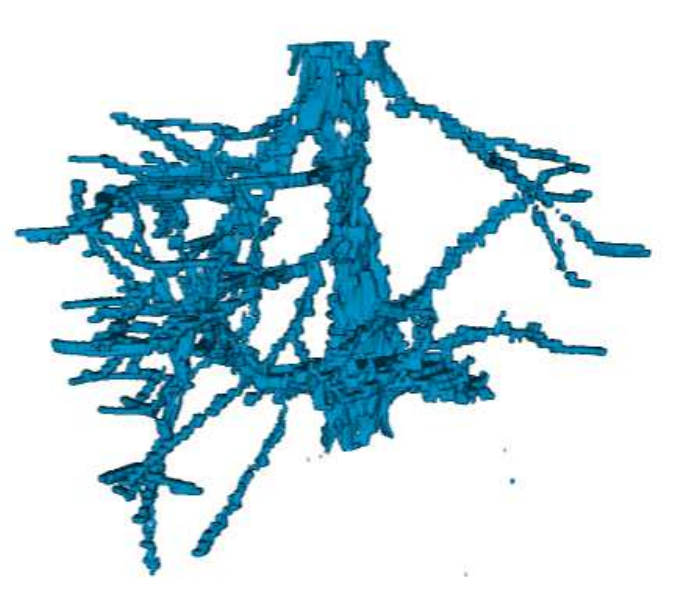} &
        \includegraphics[width=0.23\linewidth]{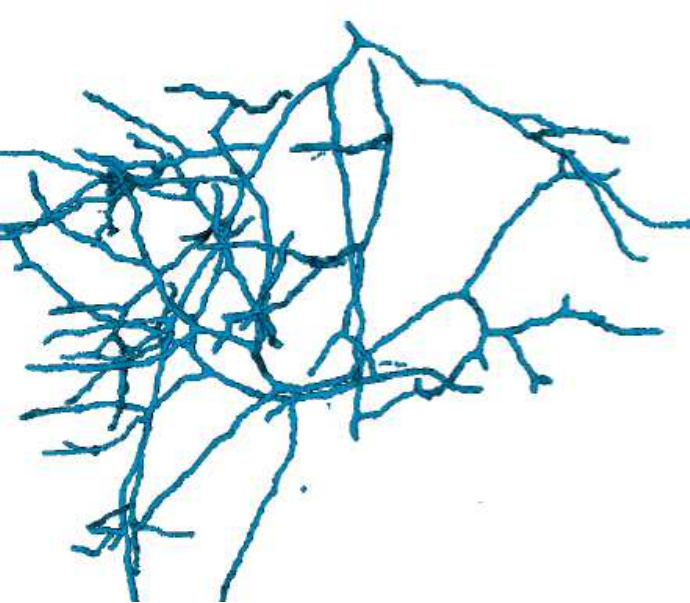} \\
            (a) Lee's skeletonization & (b) Soft-skel & (c) Soft-persistent-skel & (d) Neural skel 
    \end{tabular}

\end{figure}

\begin{table}[h]
    \centering
    \caption{Comparison of skeletonization methods with respect to the output of Lee's alogrithm. Format : Mean (standard deviation)}
    \label{tab:skel}
    \begin{tabular}{|c|c|c|c|c|}
        \hline
        \textbf{Method} & \textbf{Dice $\uparrow$(\%)} & \textbf{CLDice $\uparrow$(\%)} & \textbf{SurfaceDice$\uparrow$(\%)} & \textbf{Hausdorff $\downarrow$(mm)}
        \\
        \hline
        Soft-skel & 23.59 (4.8) & 32.70 (7.8) & 86 (7.8) & 19.19 (7.5) \\
        Soft-persistent-skel & 20.15 (4.9) & 40.11 (8.2) & 81.8 (9.9) & 18.5 (7.6) \\
        NeuralSkel (ours)  & \textbf{34.3 (6.4)} & \textbf{56.1 (6.1)} & \textbf{98.2 (4.1)} & \textbf{14 (6.37)}\\
        \hline
    \end{tabular}
    
\end{table}

\section{Results}
\subsection{DL techniques for tree skeletonization and segmentation}
Both models'(skeletonization and segmentation) results are evaluated with the Dice, ClDice, Surface Dice \cite{SurfaceD} and Hausdorff distance metrics, after taking into account the CT sampling. The tolerable threshold for the surface Dice is fixed at 2mm. The skeletonization neural network achieves a  score  of 98\% for SurfaceDice (see Table 1). 


When using it in the segmentation neural network's ClDice module, there is an improvement from the baseline. However it is not that different from using the base skeletonization functions.

\subsection{Separation algorithm to distinguish the portal and hepatic trees}

We applied the automatic separation algorithm on a 62-case dataset that was previously annotated in house with the help of liver surgeons, and an initial automatic segmentation. These 62 cases are a subset of the LiTS dataset. The surgeons have classified 35 segmentations as flawless and 18 as containing a small mixup between the two trees. The 9 remaining presented larger errors due to the heavy interconnections of the original segmentation.

After this validation, the minor issues were quickly fixed, showcasing the reduced annotation burden, and lead to 53 final segmentation volumes in total. Note that some of the initial segmentations we ran the algorithm on, had annotations that featured the main branches only.
Available cases are now 20 from IRCAD annotations with both trees, and 53 from the annotated and separated LiTS annotation. This new dataset is called LIRCAD. Training on LIRCAD improves all of our previous benchmarks across all metrics by 2 to 3 percentage points and reduces Hausdorff distance (Table \ref{table:network}).

\subsection{Automatic vessel anatomical labeling \& morphological analysis}

We now analyze the LIRCAD dataset of 77 cases, by extracting the following properties (Fig.5), showing differences between right and left livers: \textbf{\textit{tree\_ID}} (ID in the dataset), 
\textbf{\textit{tree\_label}} (portal or hepatic), 
\textbf{\textit{path\_label}} (path\_ID in the tree)
\textbf{\textit{path\_name}} (Main\_branch, couinaud\_sector,etc..) , 
\textbf{\textit{path\_length}} (in mm),
\textbf{\textit{path\_radius}} (in mm), 
\textbf{\textit{path\_ratio}} (length/radius), 
\textbf{\textit{children\_id}} , 
\textbf{\textit{children\_length}}, 
\textbf{\textit{children\_radii}},
\textbf{\textit{children\_ratio}}, 
\textbf{\textit{emergence\_angle}} (°), 
\textbf{\textit{end\_angle}} (°), 
\textbf{\textit{Power\_law\_index}}, 
\textbf{\textit{gen}} (generation of the branch), 
\textbf{\textit{n\_desc}} (number of descendants of the branch).

\begin{figure}
    \centering
    \caption{Annotation. Visualizations by Slicer 3D (www.slicer.org) \cite{slicer}}
    \label{fig:annotation}
    \begin{tabular}{ccc}
        \includegraphics[width=0.3\linewidth]{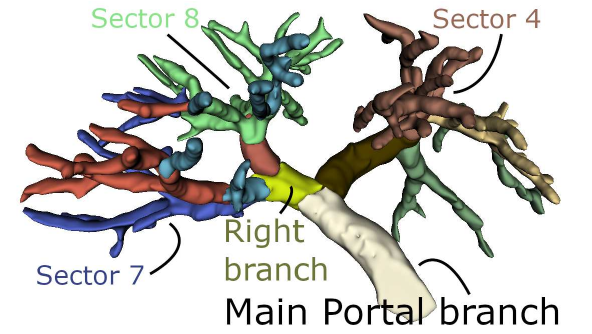} &
        \includegraphics[width=0.3\linewidth]{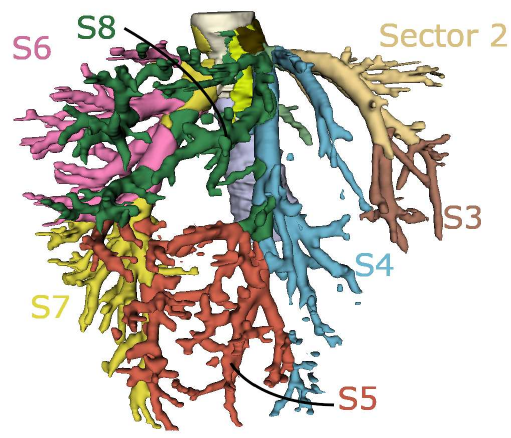} &
        \includegraphics[width=0.3\linewidth]{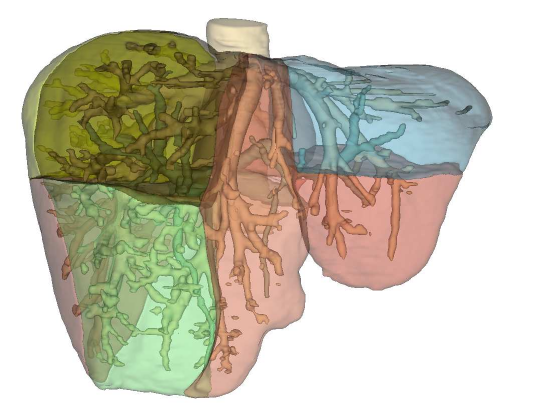} \\
        (a) Annotated portal tree & (b) Annotated hepatic tree  & (c) Couinaud segments \& vessels  
    \end{tabular}
\end{figure}
\subsubsection{Automatic labeling evaluation}
Branches of interest to the surgeons are extracted such as right portal branch and left portal branch (Fig.\ref{fig:annotation}).

\begin{figure}
    \centering
    \caption{Morphometry - portal tree properties}
    \label{fig:morphometry}
    \renewcommand{\arraystretch}{0.1} 
    \begin{tabular}{ccc}
        \includegraphics[width=0.33\linewidth]{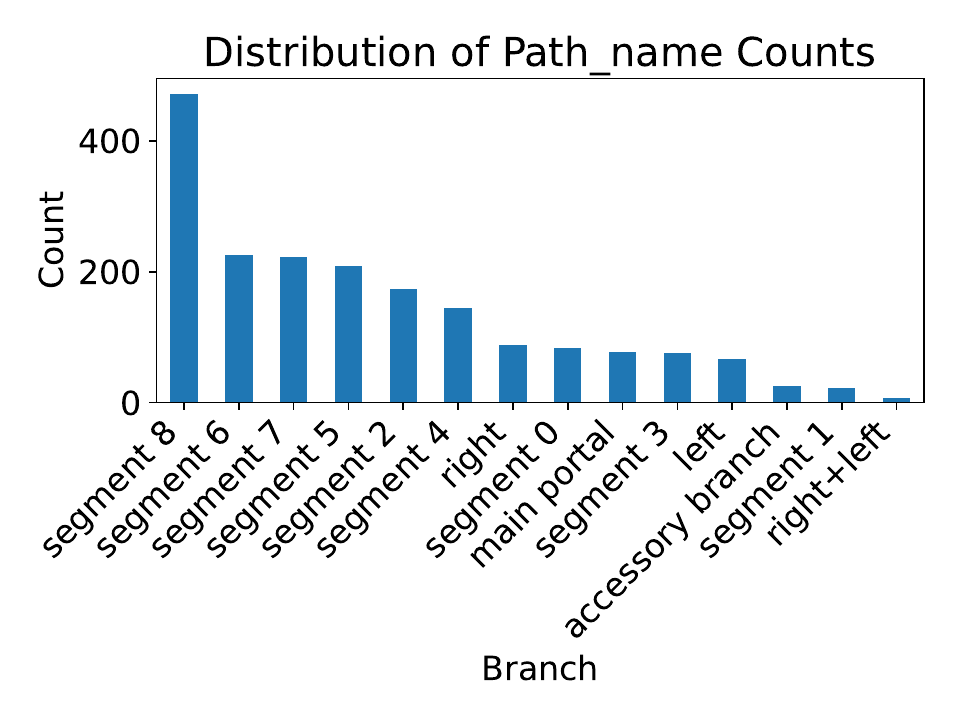} &
        \includegraphics[width=0.33\linewidth]{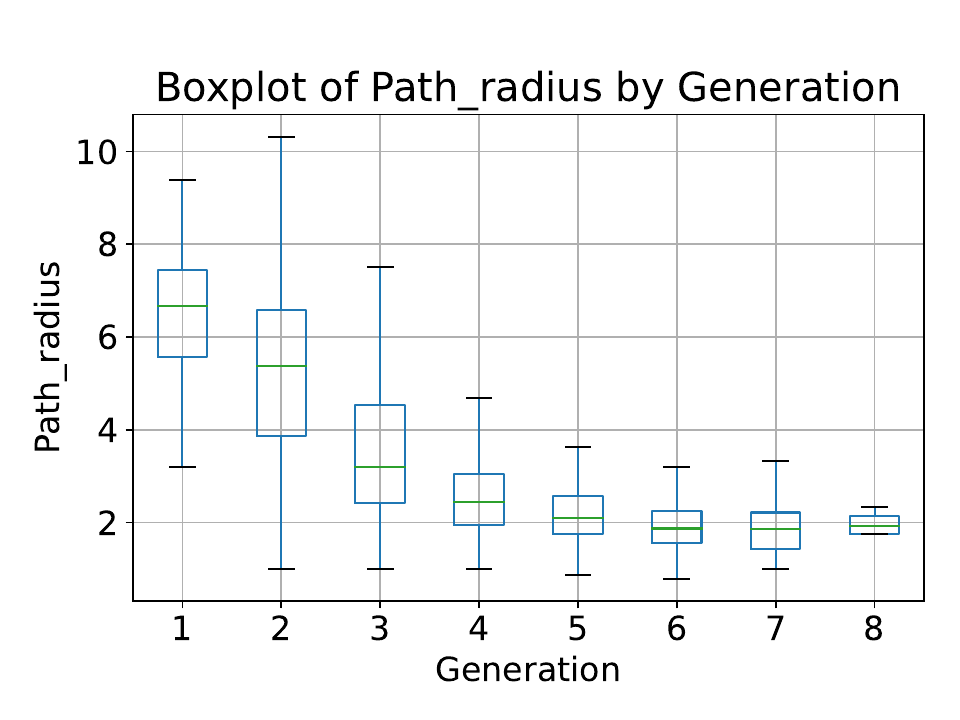} &
        \includegraphics[width=0.33\linewidth]{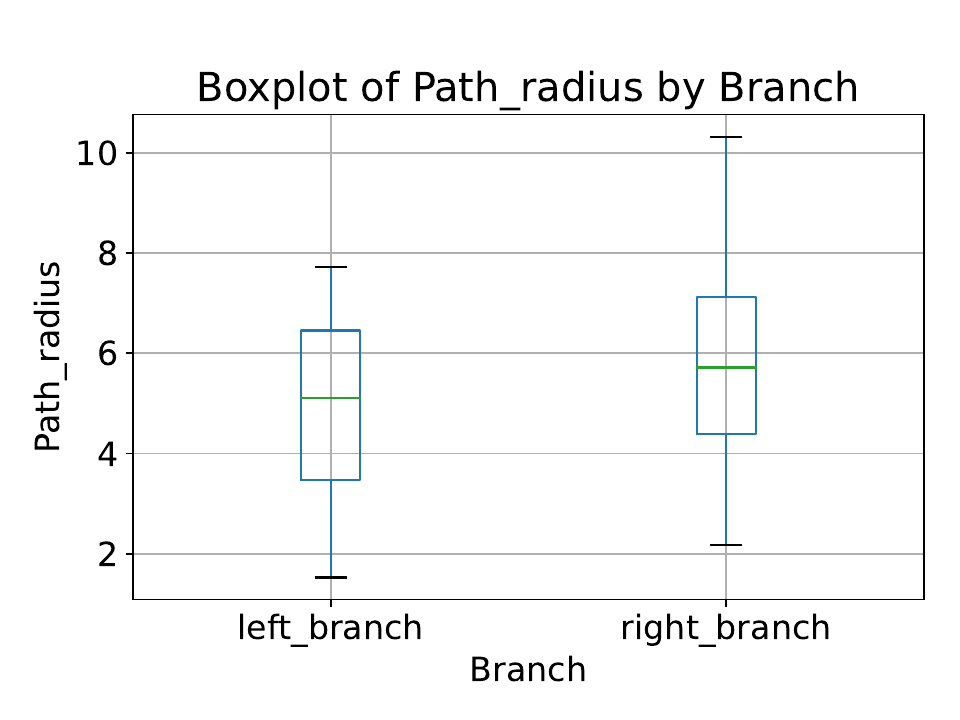} \\
    \end{tabular}
\end{figure}

\section{Discussion and Conclusion}
\textbf{Discussion}
The results show that the differentiable skeletonization performance was lackluster compared to the widely used but non-differentiable version \cite{lee}. The skeletonization neural network achieves much better centerline representations than the morphological operations as shown in Table \ref{tab:skel} and Fig.\ref{fig:skel}. Moreover, we investigated whether skeleton quality might improve ClDice results. Adding a neural network inference instead of a skeletonization algorithm consisting of a series of morphological operations does double the training time (from 80s to 160s per epoch) but it provides a more accurate and homogeneous skeleton for the ClDice computation. However, the overall segmentation performance does not improve as shown in Table \ref{table:network}. This result holds even when training for a larger number of epochs. The pretrained aspect of this neural network can make it multipurpose and is its biggest advantage. Since its input is a segmentation volume, it can theoretically be a general skeletonization algorithm for any tubular structure (such as other organs' vessels or tree roots for example).
\begin{table}[h]
    \centering
    \caption{Metrics Benchmark. Baseline: Dice + CrossEntropy. All "Dice" scores are in $\uparrow$\%, Hausdorff in mm($\downarrow$)}
    \label{table:network}
    \begin{tabular}{|c|c|c|c|c|}
        \hline
        \textbf{Model} & \textbf{Dice } & \textbf{CLDice } & \textbf{SurfaceDice} & \textbf{Hausdorff}\\
        \hline
        Baseline IRCAD & 72.28 (7.2) & 72.97 (7.8) & 79.06 (7.7) & 68.83 (53.36) \\
        CLDice IRCAD & 73.29 (6.4) & 75.24 (6.9) & 80.57 (6.6) & 65.05 (52.09)\\
        Baseline \textbf{L}IRCAD & \textbf{76.22 (5.9)}  & 77.29 (5.4) & 83.57 (5) & 68 (54.7) \\
        CLDice \textbf{L}IRCAD & 76.05 (5.2) & \textbf{77.95 (5.1)} & \textbf{83.62 (5.1)} & \textbf{61 (51.7)} \\
        NeuralSkel \textbf{L}IRCAD & 75.5 (6.0) & 76.75 (5.8) & 82.4 (6.0) & 70 (53.6) \\
        \hline
    \end{tabular}
\end{table}

The label separation algorithm works well in general, but has its corner cases: mainly highly dense and interconnected areas that should not be, due to a lackluster initial 1-label segmentation and CT Scan resolution, showcased in \textit{12} in Fig.\ref{fig:separation}. Since it occurs mostly in the distal parts of the tree, its impact for clinical application is limited. We also note that Lee skeletonization and skan are not perfect and can lead to undesirable results. They are the backbone of the algorithm and can be further optimised. However, the dataset this labelling algorithm creates, is the source of the biggest improvement of our vessel segmentation models, and opens the door for our morphological study (Fig.\ref{fig:morphometry}).

The automatic anatomical branch labeling and subsequently, the morphometry extraction, also depend on the performance of the skeletonization and sknw library which can lead to the creation of loops or big lumps of voxels. The angles obtained are limited by the fact that we are in a voxel environment. But we hope that these results can aid in tree modeling and extract viable biomarkers for surgeons and scientists.

\textbf{Conclusions}
In this paper we showcase a way to train a differentiable and precise skeletonization algorithm, a deep dive of ClDice on liver vessels which feature a 3D structure, with an extensive network of vessels that have vastly different radii.
We also present a method to obtain multi-label segmentations. This makes the annotation burden lighter as we do not need to pay attention to which tree a vessel tree belongs to, and automatically label the trees after.

Finally we also introduce a novel method to automatically label the main clinically relevant branches for surgeons during their assessment, such as the main branches and their immediate daughter branches. We also propose a labeling of the smaller branches according to their generation or Couinaud region. Finally, we extract the first computational morphometric study on human liver vessels as far as we are aware.

\textbf{Code and data}
The code, the newly obtained dataset of ``LIRCAD'' featuring 77 cases, and anatomical labels are to be published with this article. You can access them here:
\begin{itemize}
    \item  \href{https://gitlab.inria.fr/simbiotx/LiverVesselSeg}{Code link}
    \item \href{https://zenodo.org/records/13897086}{Data link}
\end{itemize}

\end{document}